\documentclass[smallextended]{svjour3}     
\smartqed  
\usepackage{graphicx}
 \usepackage{mathptmx}      
\usepackage{multirow}
\usepackage{amsmath}
\usepackage{amssymb}
\usepackage[round]{natbib}
\usepackage{epsfig}
\usepackage{psfig}
\newcommand{\1}{1 \hspace*{-0.4ex}\rule{0.10ex}{1.5ex}\hspace{0.4ex}} 

 \journalname{Advances in Statistical Analysis}
\begin{document}

\title{Latin hypercube sampling with inequality constraints}

\titlerunning{Constrained Latin hypercube sampling}        

\author{Matthieu Petelet \and
        Bertrand Iooss\footnote{Corresponding author}       \and
        Olivier Asserin        \and
        Alexandre Loredo
}


\institute{O. Asserin \at
                        CEA, DEN, DM2S, SEMT, LTA, F-91191, Gif sur Yvette, France 
\and
            B. Iooss \at
              EDF, R\&D, 6 Quai Watier, F-78401 Chatou, France \\
              Tel.: +33-1-30877969\\
              Fax: +33-1-30878213\\
              \email{bertrand.iooss@edf.fr}           
\and
           A. Loredo \at
              LRMA, EA 1859, Universit\'e de Bourgogne, France
\and
           M. Petelet \at
                        CEA, DEN, DM2S, SEMT, LTA, F-91191, Gif sur Yvette, France 
}

\date{Received: date / Accepted: date}

\maketitle

\begin{abstract}
In some studies requiring predictive and CPU-time consuming numerical models, the sampling design of the model input variables has to be chosen with caution.
For this purpose, Latin hypercube sampling has a long history and has shown its robustness capabilities.
In this paper we propose and discuss a new algorithm to build a Latin hypercube sample (LHS) taking into account inequality constraints between the sampled variables.
This technique, called constrained Latin hypercube sampling (cLHS), consists in doing permutations on an initial LHS to honor the desired monotonic constraints.
The relevance of this approach is shown on a real example concerning the numerical welding simulation, where the inequality constraints are caused by the physical decreasing of some material properties in function of the temperature.

\end{abstract}

\section{Introduction}

With the advent of computing technology and numerical methods, investigation of computer code experiments remains an important challenge.
Complex computer models calculate several output values which can depend on a large number of input parameters and physical variables.
These computer models are used to make simulations as well as predictions, uncertainty and sensitivity analyses or to solve optimization problems \citep{fanli06,kle08,derdev08,levste10}.

However, complex computer codes are often too time expensive to be directly used to perform such studies.
For uncertainty propagation and sensitivity analyses, it has been shown that the sampling design is one of the key issues \citep{salcha00,fanli06}.
Moreover, to avoid the problem of huge calculation time, it is often useful to replace the computer code by a mathematical approximation, called a surrogate model or a metamodel \citep{simpep01,fanli06,volioo08}.
The optimal exploration of the variation domain of the input variables is therefore especially important in order to avoid non-informative simulation points \citep{sob76,simlin01,burste06a,ioobou10,levste10}.

Thirty years ago, \citet{mckbec79} have introduced the concept of Latin hypercube sampling (LHS) for numerical experiments.
Compared to simple random sampling (SRS) which insured independence between samples, LHS ensures the full coverage of the range of the input variables.
More precisely, LHS allows to accurately reproduce the one-dimensional projections of the input sampling design.
In terms of uncertainty and sensitivity analyses, it has been theoretically and experimentally proved that LHS is more precise and robust than simple random sample \citep{ste87,owe92,salcha00,heldav03}. 
Moreover, in the last twenty years, several improvements have been proposed in order to optimize the space filling properties of LHS designs \citep{par94,fanli06,pisvic10,joufra10}.

Our starting point is that the initial LHS algorithm supposes independence between input variables while in some situations, this assumption is irrelevant.
First, let us recall the two forms of dependencies between variables: the statistical and the physical ones.
\begin{itemize}
\item
The correlation coefficient is the simplest measure of the statistical dependence between two variables.
For the SRS, the so-called joint normal transform method consists in inducing a correlation structure on the transformed marginals \citep{kurcoo06}.
The rank correlation coefficient, based on the rank transformation (which turns each variable value to its rank in the sample), is known as a more robust measure.
\citet{imacon82} have introduced an algorithm to consider rank correlations between variables in LHS.
Some limitations of these two dependence measures have led to the introduction of other statistical dependence modeling (copulae, vines, etc., see \citet{kurcoo06}).
\item
Physical dependencies between variables can arise when a variable has a formal relation in function of other variables.
Such input constraints have been studied by \citet{bor08} which has proposed a novel way to solve the sensitivity analysis problem in presence of equality constraints.
Another currently encountered physical dependence, which is the subject of this paper, concerns the existence of inequality relations between the variables.
It is the case when one variable is physically constrained to be larger (respectively smaller) than another.
For example, a geometric parameter (radius, height, etc.) of two physical objects can be subject to a rigorous increasing order if one object is included inside the other.
\end{itemize}

When building the sampling design, the inequality constraints have to be honored in order to avoid some physical incoherence in the input sets that will be run with the computer model.
A first solution could be to sequentially simulate the input variables, allowing to bound one variable by another in order to enforce an inequality constraint.
However, as we will see, this procedure affects the one-dimensional projections of the sampling design.
We need a procedure which separates the effect of dependence (the inequality constraints) from the effects of marginal distributions (i.e. the probability laws defined for each variable). 
To attain this objective, we propose an algorithm which builds a LHS satisfying the inequality constraints.

This paper is devoted to the detailed presentation of this algorithm, called the constrained LHS (cLHS).
In the next section, we introduce this algorithm by giving some examples.
We compare it with a SRS-based algorithm and illustrate the algorithmic performances.
In the third section, we explain in details the cLHS algorithm.
As the inequality constraints can be too stringent to find a cLHS, we derive a necessary and sufficient condition proving its existence from an initial LHS.
Then, our methodology is applied on a real problem involving welding simulation models.
A conclusion gives finally some prospects to improve the cLHS algorithm.


\section{The sampling techniques}\label{par2} 

The goal of the sampling step is to generate a matrix $\mathbf{X}^n=(x_j^{(i)})_{i=1..n,j=1..p}$, where $n$ is the number of experiments and $p$ is the number of variables. 
The most common sampling method is indisputably the pure Monte Carlo (i.e. SRS), mainly because of its simplicity \citep{gen03}. It consists of randomly sampling $n$ independent input variables. 
However, it is known to have poor space filling properties:
 SRS leaves large unsampled regions and can propose too close points. 
An example of a SRS is presented on Figure~\ref{fig:comparaisonSRSLHS} (a).

\subsection{Latin hypercube sampling}

\begin{figure}[!ht]
$$\psfig{figure=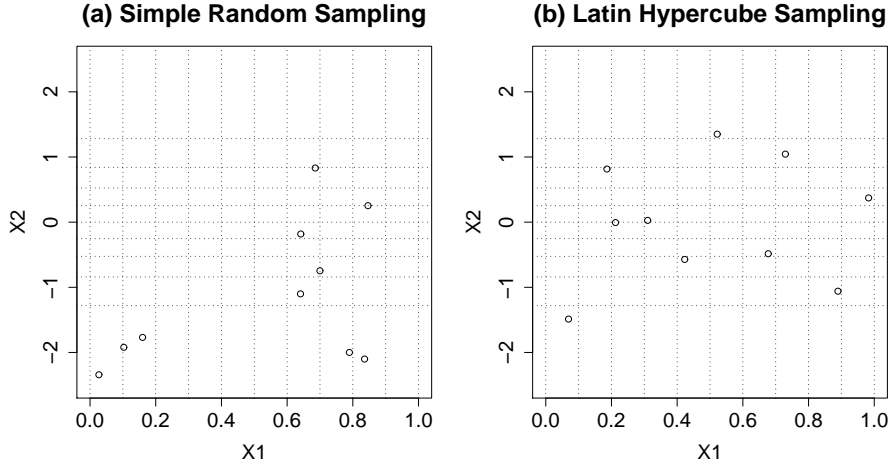,width=0.7\columnwidth,angle=-90}$$

\vspace{-1cm}
\caption{Examples of two ways to generate a sample of size $n=10$ from two variables $\mathbf{X} = [X_1,X_2]$ where $X_1$ has a uniform distribution ${\cal U}[0,1]$ and $X_2$ has a normal distribution ${\cal N}(0,1)$.}
      \label{fig:comparaisonSRSLHS}
\end{figure}

\citet{mckbec79} suggested an alternative method of generating $\mathbf{X}^n$ that they called Latin hypercube sampling (LHS) which is an extension of stratified sampling. LHS ensures that each of the input variables has all of its range represented.
Let the range of each variable $X_j$, $j = 1\ldots p$, be simultaneously partitioned into $n$ equally probable intervals.
We note $X_j^n$ the $n$-sample of the variable $X_j$.
A LHS of size $n$ is obtained from a random selection of $n$ values --- one per stratum --- for each $X_j$. Thus we obtain $p$ $n$-tuples that form the $p$ columns of the $n \times p$ matrix of experiments $\mathbf{X}^n$ generated by LHS: the $i^{th}$ line of this matrix contains the $p$ input variables and will correspond to the $i^{th}$ code execution. 
Once a point is selected in an interval, no other point could be selected in this interval (see Figure~\ref{fig:comparaisonSRSLHS} (b)).
Let us remark that the partition into equally probable intervals allows to take into account non uniform densities of probability like a normal distribution for example. 
Figure ~\ref{fig:comparaisonSRSLHS} shows $10$ samples of two random variables obtained with SRS and LHS schemes.
We can see that the result of LHS is more spread out and does not display the clustering effects found in SRS.

Mathematically, if $X_1$, \ldots, $X_p$ are mutually independent random variables with invertible continuous distribution functions $F_j$, $j=1,\ldots,p$, respectively, then the LHS $i$-th sample for the $j$-th variable can be created as
\begin{equation}
x_j^{(i)} = F_j^{-1}\left( \frac{\pi_j^{(i)}-\xi_j^{(i)}}{n}\right) \;,
\end{equation}
where the $\pi_{j}$ are independent uniform random permutations of the integers $\{1,2,\ldots,n\}$, and the $\xi_j^{(i)}$ are independent ${\cal U}[0,1]$ random numbers independent of the $\pi_{j}$.

\vspace{-1.5cm}

\vspace{-0.5cm}
\subsection{Constrained simple random sampling}

To take into account inequality constraints between variables, the simplest approach is based on SRS and consists on bounding one variable by another in order to enforce the inequality constraints.
This approach is called the constrained Simple Random Sampling (cSRS).

\begin{figure}[!ht]
$$\hspace{-1cm}\psfig{figure=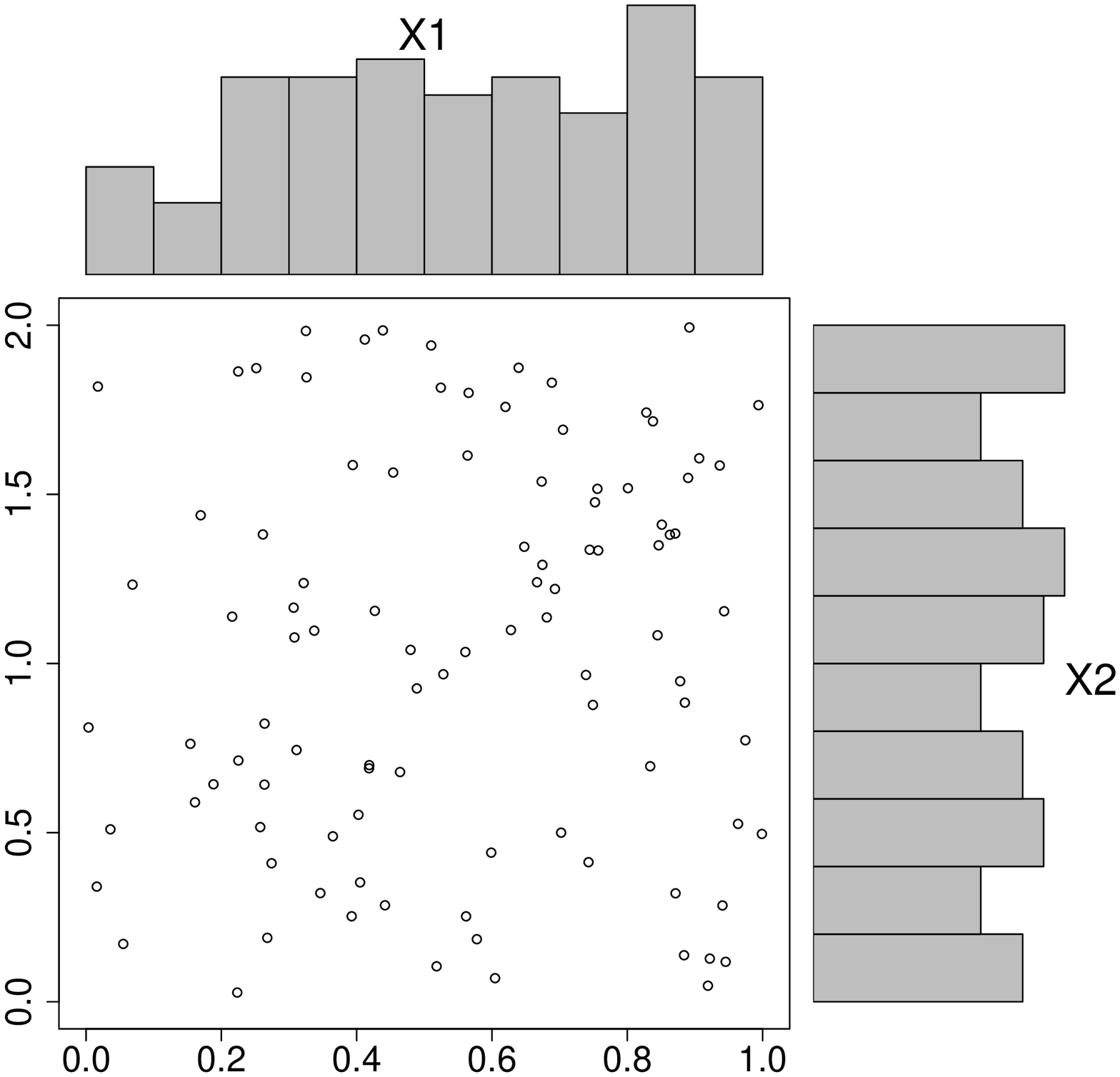,width=0.45\columnwidth,angle=-90}
\hspace{-1.5cm} \psfig{figure=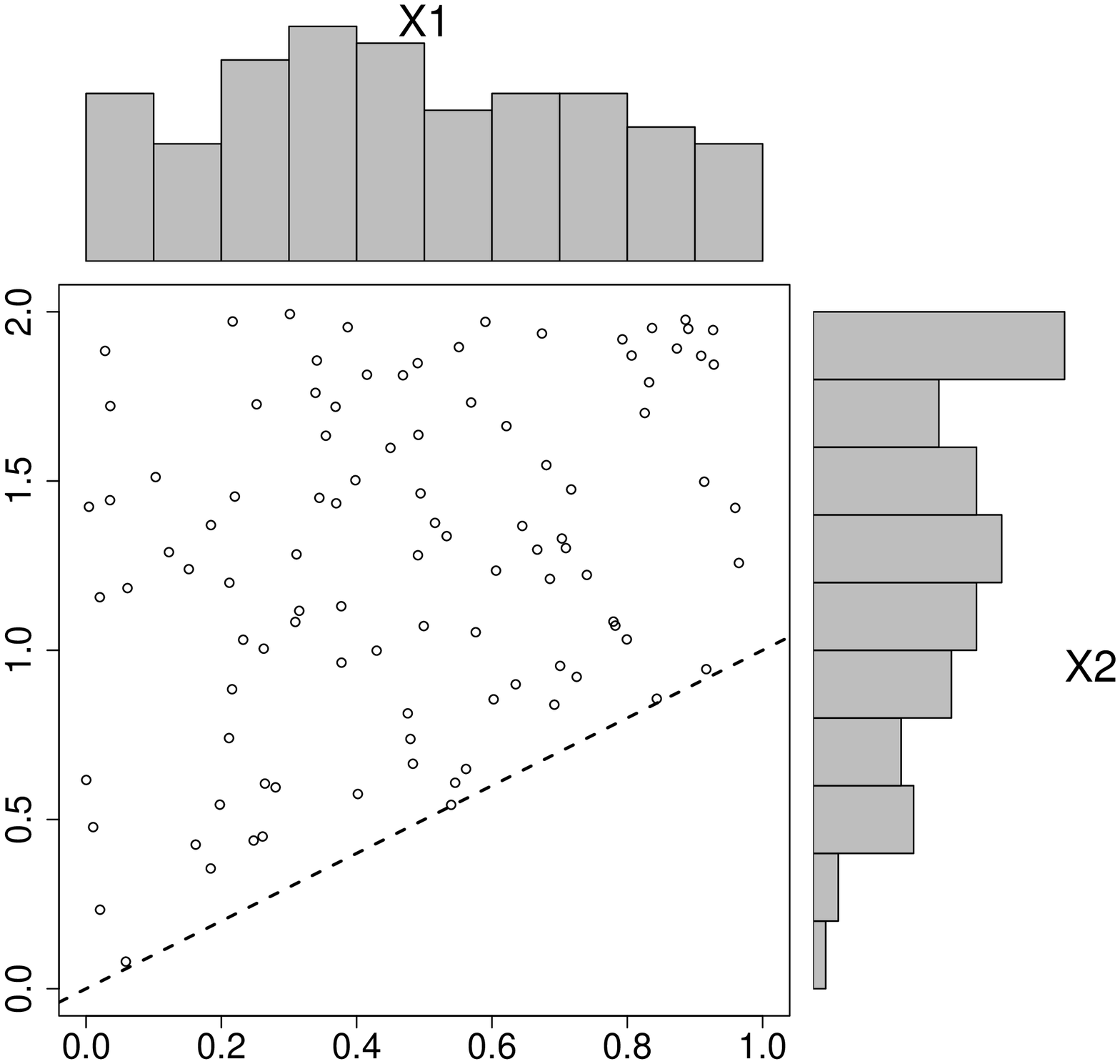,width=0.45\columnwidth,angle=-90}$$

\vspace{-0.2cm}
\centerline{(a) Simple random sampling (SRS) \hspace{1.cm} (b) Constrained simple random sampling (cSRS)}

\caption{Comparison between Monte Carlo samples of size $n=100$ from two variables $\mathbf{X} = [X_1,X_2]$ where $X_1 \sim {\cal U}[0,1]$ and $X_2 \sim {\cal U}[0,2]$. The inequality constraint for cSRS is $X_1<X_2$.}
      \label{fig:comparaisonSRScSRS}
\end{figure}

In Figure \ref{fig:comparaisonSRScSRS}, we see the effect of an inequality constraint between two variables in terms of bivariate plots.
Of course, the introduction of the truncation during the simulation creates statistical dependences between $X_1^n$ and $X_2^n$.
This correlation depends on the distribution functions of $X_1$ (called $F_1$) and $X_2$ (called $F_2$).
In our example, the correlation coefficient $\rho(X_1,X_2)$ is worth $31\%$.
In Figure \ref{fig:comparaisonSRScSRS}, the one-dimensional marginal projections of the samples are also shown.
For the cSRS, the one-dimensional marginal of $X_2^n$ does not correspond to $F_2$ anymore but to a transformed distribution $F'_2$ (which depends on $F_1$).

\vspace{-0.5cm}

\begin{figure}[!ht]
$$\psfig{figure=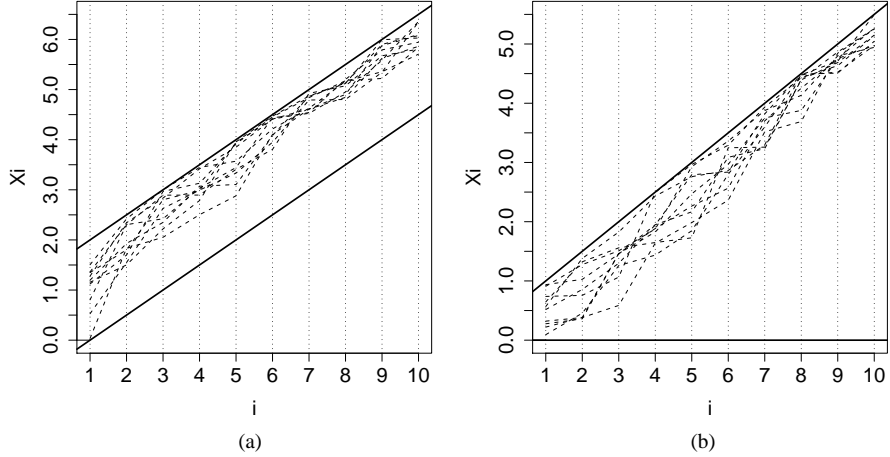,width=0.7\columnwidth,angle=-90}$$

\vspace{-1.2cm}
\hspace{3cm}(a) \hspace{5.5cm} (b)
\caption{Constrained simple random samples of size $n=10$ from $p=10$ variables $X_i$, $i=1,\ldots,10$, with $X_i < X_{i+1}$ for $i=1,\ldots,9$. The upper and the lower curves represent the bounds of the variation ranges for these $10$ variables. (a) $X_i \sim {\cal U}[0+\frac{i-1}{2},2+\frac{i-1}{2}]$; (b) $X_i \sim {\cal U}[0,1+\frac{i-1}{2}]$. }
      \label{fig:comparcSRS_10}
\end{figure}

This problem becomes more dramatic when the input dimension increases and when several sequential inequality constraints have to be satisfied, as for example if $X_i < X_{i+1}$ for $i=1,\ldots,p-1$ and $p$ is large.
Figure~\ref{fig:comparcSRS_10}~(a) shows an example of cSRS of $p=10$ variables with such increasing constraints.
In this graph, each observation is represented as a line.
Because of the sequential algorithm starting at $X_1$, all the curves are concentrated near the upper bound curve of the variables.
Figure~\ref{fig:comparcSRS_10}~(b) clearly reveals that the sampling of the first variable $X_1$ is adequate with its uniform distribution, and that the samples of the following variables ($X_2$ to $X_{10}$) progressively 
take place in the upper region of their variation range.


In summary, in some practical situations, users would like to simulate samples which follow all one-dimensional marginals and which take into account some inequality constraints.
The following section proposes such an algorithm. 

\subsection{Constrained Latin hypercube sampling}

In order to follow all one-dimensional marginals, our sampling procedure uses LHS.
Our method, first proposed in \cite{pet07}, consists in doing permutations on an initial LHS to enforce the desired monotonic constraint. It is based on the fact that permuting two values of a variable in a LHS does not break the LHS structure of the sample \citep{imacon82}. 
An appropriate algorithm scans the starting LHS to find the couples of values that violate the monotonic constraint. 
Then the algorithm finds and executes the combinations of permutations which have to be done
to satisfy the inequality constraint between $X_i$ and $X_{i+1}$ for the $n$ experiments.
Details of the algorithm are given in section \ref{sec:algo}.

\begin{figure}[!ht]
$$\hspace{-1cm}\psfig{figure=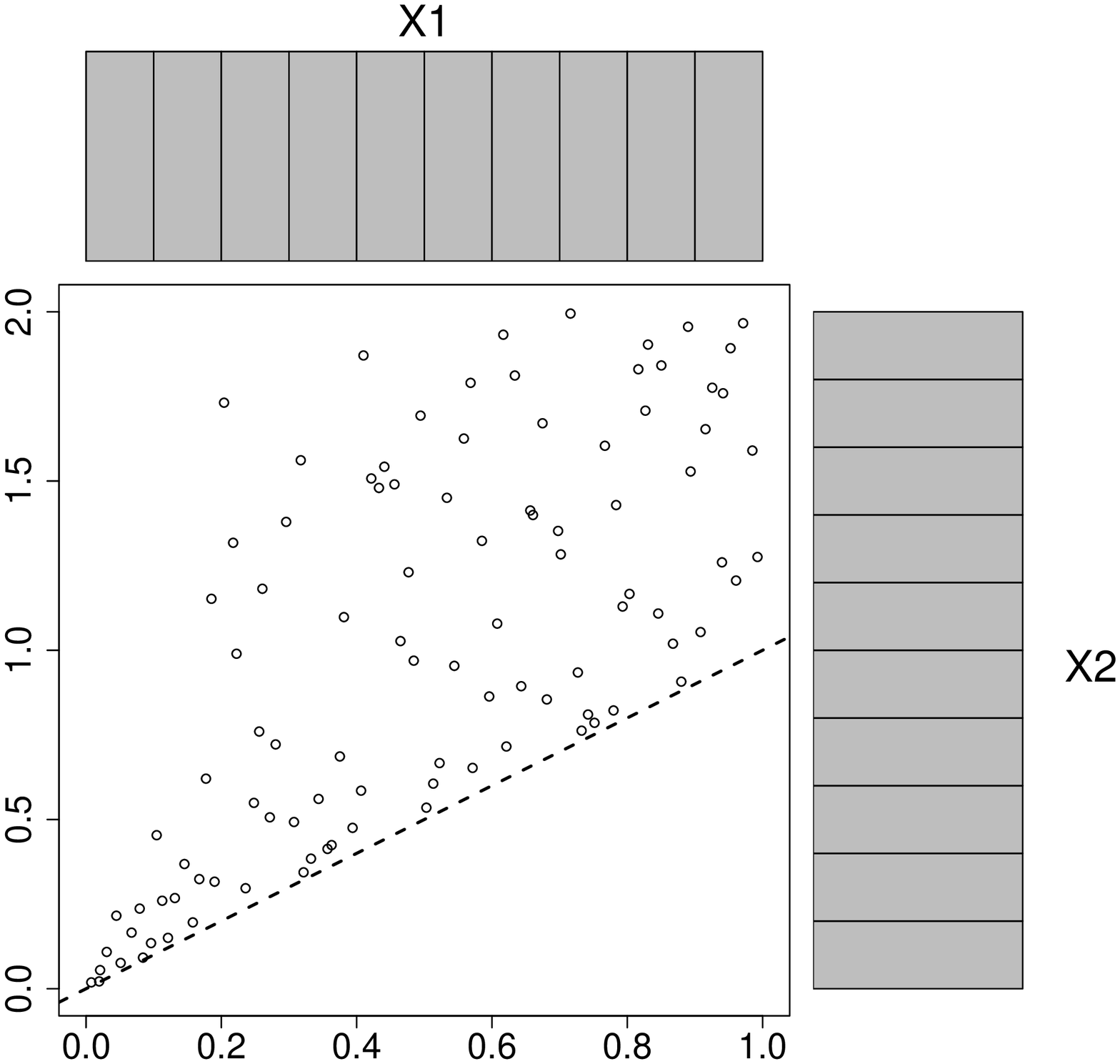,width=0.45\columnwidth,angle=-90}
\hspace{-1.5cm} \psfig{figure=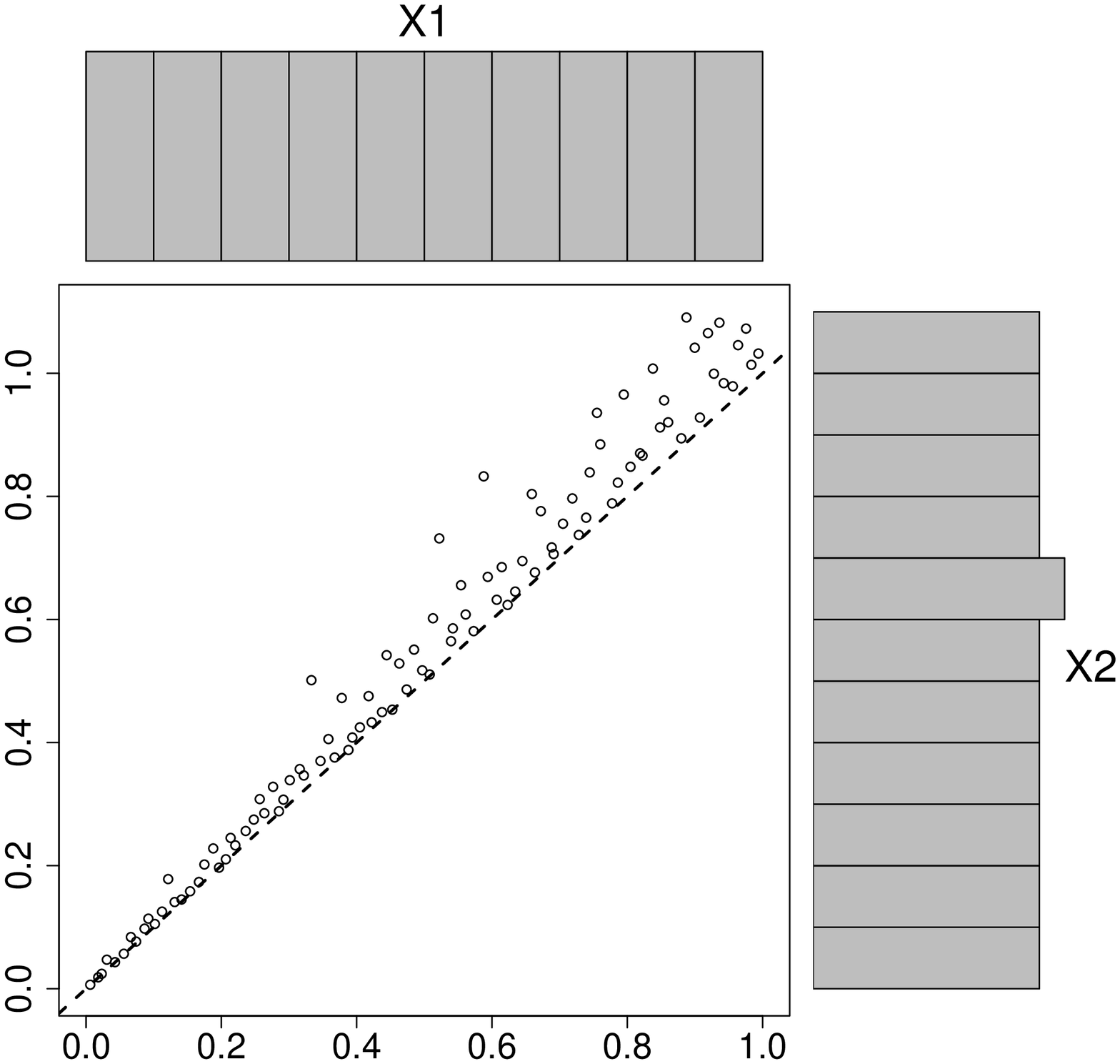,width=0.45\columnwidth,angle=-90}$$

\vspace{-0.2cm}
\hspace{2.5cm}(a) \hspace{5.5cm} (b)

\caption{Comparisons between constrained Latin hypercube samples of size $n=100$ from two variables $\mathbf{X} = [X_1,X_2]$ with $X_1 \sim {\cal U}[0,1]$ and the inequality constraint
 $X_1<X_2$: (a) $X_2 \sim {\cal U}[0,2]$; (b) $X_2 \sim {\cal U}[0,1.1]$.}
      \label{fig:comparaisoncLHS}
\end{figure}

Figure~\ref{fig:comparaisoncLHS} shows the work done on a couple of parameters on which an increasing constraint is enforced.
The distribution of $X_1$ and $X_2$ are kept uniform (the slight variation of the height of one class in the histogram of X2 is not significant). 
In the bivariate plots, we see that the cLHS constraints (the increasing inequality and the honoring of all one-dimensional marginals) tend to gather the sample points along the inequality frontier line $X_1=X_2$.
It appears that the severity of the inequality constraint effects strongly depend on the one-dimensional marginal distributions $F_1$ and $F_2$ of the variables.
The limit case is illustrated on Figure~\ref{fig:comparaisoncLHS}~(b) where $X_1$ and $X_2$ have the same one-dimensional marginal distributions $F_1$ and $F_2$ (then the same upper and lower bounds).
In such a case where half the area of the bivariate plot is forbidden all the points are located on this frontier line.
This effect results of a too severe constraint and reveals the need of a constraint intensity measurement.
\cite{pet07} has defined this constraint intensity measurement as the ratio between the triangular forbidden area ($S_T$) in the bivariate plot and the rectangular area ($S_R$) of the domain defined by the upper and lower bounds of the variables:
    \begin{equation}\label{eq:gamma}
        \gamma=\frac{S_T}{S_R}\;.
    \end{equation}

The $\gamma$ measurement can be used if each of the input variable has some upper and lower bounds, i.e. if the support of their distribution function is defined on a bounded domain.
In the general case of an inequality constraint between $X_i$ and $X_j$, respectively defined on $[b_i,h_i]$ and $[b_j,h_j]$, we obtain
    \begin{equation}\label{eq:gammaij}
\gamma(X_i,X_j)=\frac{S_T(X_i,X_j)}{S_R(X_i,X_j)} = 
\left\{\begin{array}{rcl}
\displaystyle \frac{(h_i-b_j)^2}{2(h_i-b_i)(h_j-b_j)} & \; \mbox{ for the constraint } \; & X_i < X_j \;,\\
\displaystyle \frac{(h_j-b_i)^2}{2(h_i-b_i)(h_j-b_j)} & \; \mbox{ for the constraint } \; & X_i > X_j \;.
\end{array}\right.
\end{equation}
The intensity constraint measurements for the Figure ~\ref{fig:comparaisoncLHS} cases are worth $\gamma(X_1,X_2)=25\%$ for (a) and $\gamma(X_1,X_2)=45.5\%$ for (b).
With some heuristic arguments, \cite{pet07} has found a nearly linear link between $\gamma(X_i,X_j)$ and the correlation coefficient $\rho(X_i,X_j)$ for $\gamma(X_i,X_j) \in [0,0.3]$:
\begin{equation}\label{eq:relgamcor}
\rho(X_i,X_j) \simeq 2.778 \;\gamma(X_i,X_j) \;.
\end{equation}
As $\gamma$ is positive, the correlation coefficient will be always positive. 
For example, this relation shows that if the inequality constraint is kept smaller than $15\%$, the correlation between the variables will be smaller than $40\%$.

\begin{figure}[!ht]
$$\psfig{figure=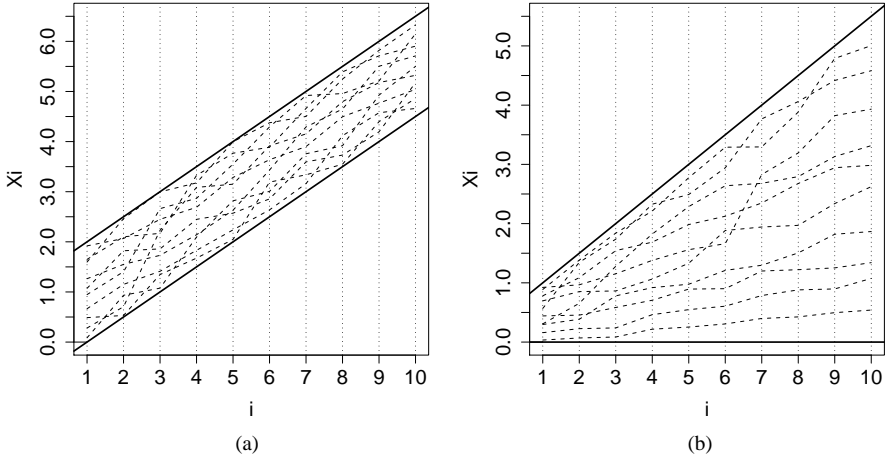,width=0.7\columnwidth,angle=-90}$$

\vspace{-1.2cm}
\hspace{3cm}(a) \hspace{5.5cm} (b)
\caption{Constrained Latin hypercube samples of size $n=10$ from $p=10$ variables $X_i$, $i=1,\ldots,10$, with $X_i < X_{i+1}$ for $i=1,\ldots,9$. The upper and the lower curves represent the bounds of the variation range for these $10$ variables. (a) $X_i \sim {\cal U}[0+\frac{i-1}{2},2+\frac{i-1}{2}]$; (b) $X_i \sim {\cal U}[0,1+\frac{i-1}{2}]$. }
      \label{fig:comparcLHS_10}
\end{figure}

For the same cases than in Figure~\ref{fig:comparcSRS_10}, Figure~\ref{fig:comparcLHS_10} shows the cLHS of $p=10$ variables with sequential increasing constraints.
As before, in this graph, each observation is represented as a line.
At present, the curves correctly fill the variation ranges of all the variables.
The constraint intensity measurements for Figure~\ref{fig:comparcLHS_10} (a) are worth $\gamma(X_i,X_{i+1})=18.75\%$ for $i=1,\ldots,9$.
This value is rather suitable: correlations between variables are smaller than $52\%$ (value obtained thanks to Eq. (\ref{eq:relgamcor})).
For Figure~\ref{fig:comparcLHS_10} (a), the constraint intensity measurements increase from $\gamma(X_1,X_2)=33.33\%$ to $\gamma(X_9,X_{10})=45.45\%$.

When the upper and lower bounds of the variables are similar, the cLHS tends to give homothetic translated trajectories, as shown by Figure~\ref{fig:comparcLHS_10}~(b).
In the limit case, if all the $X_i$'s have the same one-dimensional marginal distributions (then the same upper and lower bounds), the obtained curves are parallel and regularly spaced between the lower bound curve and the upper bound curve.
This is one of the drawback of our algorithm, caused by the imposed LHS structure.
Moreover, the feasability of the cLHS depends on the bound values of the constrained variables.
For example, for the constraint $X_i<X_j$, the algorithm does not work if $h_i > h_j$ or if $b_i > b_j$.
From Eq. (\ref{eq:gammaij}), this implies that the constraint intensity measurement $\gamma(X_i,X_j)$ is upper bounded by $0.5$.

\vspace{-2.5cm}

\vspace{-1cm}
\section{The constrained Latin hypercube sampling algorithm}\label{sec:algo}



In the following, we explain the cLHS algorithm in the case of a strict increasing constraint between two variables $X_1$ and $X_2$ (with distribution functions $F_1$ and $F_2$ respectively).
The developments for the strict decreasing constraint case are exactly the same, by inverting the inequalities sense.
Moreover, the extension of our algorithm to non strict inequality constraints is straightforward.

For the increasing constraint case, we assume the following hypotheses:
\begin{itemize}
\item $\mathbf{X}=(X_1,X_2)$ is defined on a bounded domain $\mathcal{X} \in \mathbb{R}^2$. The support of $F_j$ for $j=\{1,2\}$ is $[b_j,h_j]$;
\item The bounds are subject to the following inequalities:
\begin{equation}\label{eq:bounds}
b_1 \leq b_2 \;\mbox{ and }\; h_1 \leq h_2 \;.
\end{equation}
These inequalities seem natural: if we impose some increasing constraint between $X_1$ and $X_2$, we hope that the same increasing constraints exist for their minimal and maximal bounds.
\end{itemize}

Let us define the matrix $\mathbf{C}^n = \mathbf{C}(X_1^n,X_2^n)$ of size $n \times n$:
\begin{equation}\label{eq:compat}
\mathbf{C}^n = \left( c_{ij} \right)_{i=1..n,j=1..n} = 
\begin{pmatrix} 
\1_{x_2^{(1)} > x_1^{(1)}} & \cdots & \1_{x_2^{(1)} > x_1^{(n)}} \\
\vdots & \ddots & \vdots \\
\1_{x_2^{(n)} > x_1^{(1)}} & \cdots & \1_{x_2^{(n)} > x_1^{(n)}}
\end{pmatrix} \;,
\end{equation}
where $\1_{x>y}=1$ if $x>y$ and $\1_{x>y}=0$ otherwise.
$\mathbf{C}^n$ is called the compatibility matrix between $X_1^n$ and $X_2^n$.
This matrix allows to identify which combinations of elements of $X_1^n$ and $X_2^n$ are incompatible, i.e. those with a decreasing relation.
Therefore, the inequality constraint between the two samples is honored if the diagonal of $\mathbf{C}^n$ contains only $1$'s.

For our cLHS algorithm, if we choose to leave $X_1^n$ unchanged and give the possibility to permute some elements of $X_2^n$, we define our final objective as getting a sample $X_2^{'n}$ such that
\begin{equation} \label{eq:objective}
\sum_{i=1}^n c'_{ii} = n \;,
\end{equation}
with $\mathbf{C}' = \left( c'_{ij} \right)_{i=1..n,j=1..n}$ the compatibility matrix between $X_1^n$ and $X_2^{'n}$.

At present, it would be convenient to know if this objective can be achieved for a specific sample $\mathbf{X}^n$.
Let us define the sample vector
\begin{equation}
\mathbf{S}^n = (S_i)_{i=1..n} = \left( \sum_{j=1}^n c_{ij} \right)_{i=1..n} \;.
\end{equation}
$S_i$ gives the number of elements of $X_1^n$ which satisfy the constraint with $x_2^{(i)}$.
We also define $\left(\tilde{S}_i\right)_{i=1..n}$ the ordered sample of $(S_i)_{i=1..n}$.
To insure that we can obtain a sample $X_2^{'n}$ (by permutations of the $X_2^n$ elements) satisfying the increasing constraint, the following assertions have to be true:
\begin{itemize}
\item The smallest element of $X_2^n$ have one or more than one smaller elements in $X_1^n$, which is equivalent to say that $\min(X_2^n) \geq \min(X_1^n)$, then to say that $\tilde{S}_1 \geq 1$;
\item \ldots
\item The $i^{\mbox{th}}$-smallest element of $X_2^n$ have $i$ or more than $i$ smaller elements in $X_1^n$, which is equivalent to say that $\tilde{S}_i \geq i$;
\item \ldots
\item The $n^{\mbox{th}}$-smallest element of $X_2^n$ have $n$ or more than $n$ smaller elements in $X_1^n$, which is equivalent to say that $\tilde{S}_n \geq n$; 
\end{itemize}

From these assertions, we obtain the following result:\\
\\
{\it
\noindent{\bf Proposition:} If $(X_1^n,X_2^n)$ is a LHS, the inequality
\begin{equation}\label{eq:criterion}
\min[ \left(\tilde{S}_1 \;\cdots\; \tilde{S}_n \right) - (1 \;\cdots\; n) ] \geq 0 
\end{equation}
is a necessary and sufficient condition to guarantee the existence of a sample $X_2^{'n}$ such that $(X_1^n,X_2^{'n})$ satisfies the increasing constraint $X_1 < X_2$, where $X_2^{'n}$ is obtained from permutations of the elements of $X_2^n$.
}
\vspace{0.5cm}

Therefore, the first step of our methodology will be to test this criterion.
If equation (\ref{eq:criterion}) is not verified for a chosen $\mathbf{X}^n=(X_1^n,X_2^n)$, a new sample for $X_2^n$ (keeping the LHS property for $\mathbf{X}^n$) is created and the scan starts again. 
Our hypotheses on the bounds of $X_1$ and $X_2$ (Eq. (\ref{eq:bounds})) guarantee that a LHS satisfying the increasing constraint exists.


We start from an initial LHS $\mathbf{X}^n=(X_1^n,X_2^n)$, with a compatibility matrix $\mathbf{C}$, satisfying the existence criterion (\ref{eq:criterion}).
We want to obtain the LHS $\mathbf{X}^{'n}=(X_1^n,X_2^{'n})$ (with a compatibility matrix $\mathbf{C}'$) satisfying the increasing constraint between $X_1$ and $X'_2$.
Our objective is therefore to obtain the result of equation (\ref{eq:objective}).

Let us note $\tilde{X}_1^n$ the reverse ordered sample of $X_1^n$. 
This vector $\tilde{X}_1^n$ contains the elements $\tilde{x}_1^{(1)} \geq \tilde{x}_1^{(2)} \geq \cdots \geq \tilde{x}_1^{(n)}$.
We put in the sample vector $A^n$ a sequence of indices: the indices in $X_1^n$ of the $\tilde{X}_1^n$ elements.
Mathematically, it follows that 
\begin{equation}
x_1^{(A_i)} = \tilde{x}_1^{(i)} \;.
\end{equation}

Our permutation algorithm is based on the treatment of the $X_1^n$ elements in a sequential manner (because the constraint is more difficult to be satisfied by the first values of $X_1^n$).
We describe the algorithm with the following four steps.

\vspace{0.5cm}
\noindent
{\it Algorithm cLHS:}
\begin{enumerate}
\item
Initialisation: $\mathbf{C}^{'n} = C^n$ and $\mathbf{B}' = (1 \;\cdots\; n)$.

\item
For $i=1\ldots n$:

\begin{itemize}
\item
We put in the vector $\mathbf{B}$ the indices in $X_2^n$ of the elements compatible with $x_1^{(A_i)}$:

\vspace{0.5cm}
\hspace{1cm} k=1

\hspace{1cm} For $j=1\ldots n$:

\hspace{1.5cm} if $c'_{jA_i}=1$ then $B_k = k$ and $k=k+1$
\vspace{0.5cm}

\item
We randomly choose an element in $\mathbf{B}$ and put it in $B'_i$.

\item
The index $B'_i$ corresponds to the one that will be permuted in $X_2^n$.
We turn to zero the line $B'_i$ in the compatibility matrix $\mathbf{C}^{'n}$ (in order to block up the index $B'_i$):

\vspace{0.5cm}
\hspace{1cm} For $j=1\ldots n:\;\; c'_{B'_i j} = 0$
\vspace{0.5cm}
\end{itemize}

\item
The vector $\mathbf{B}'$ contains the indices that will be used to make the permutations in $X_2^n$.
We obtain the new sample of the variable $X_2$:

\begin{equation}
\left( X'_2 \right)_{i=1..n} = \left( X_2 \right)_{i=B'_1..B'_n} \;.
\end{equation}

\item
Finally, the permutation matrix $\mathbf{C}^{'n}$ is calculated with $X_1^n$ and $X_2^{'n}$ by equation (\ref{eq:compat}) in order to test the equality of equation (\ref{eq:objective}). 

\end{enumerate}

\noindent
{\it End of algorithm}
\vspace{0.5cm}

The extension of the CLHS algorithm to the multivariate case $\mathbf{X}=(X_1,\ldots,X_p)$ with $X_j < X_{j+1}$ for $j=1,\ldots,p-1$ is straightforward and is done in a sequential manner.
We first simulate a LHS $\mathbf{X}^n=(X_1^n,\ldots,X_p^n)$.
Leaving $X_1^n$ unchanged, we sequentially build with the cLHS algorithm $X_j^{'n}$ from $X_{j-1}^{'n}$ and $X_j^n$ for $j=2,\ldots,p$.
At each step $j$, before applying the algorithm, the criterion (\ref{eq:criterion}) is tested.
If this criterion is not verified, a new LHS for $X_j^n$ is created, and so on until the criterion is verified.

\section{The cLHS algorithm in an example}

\begin{figure}[!ht]
$$\psfig{figure=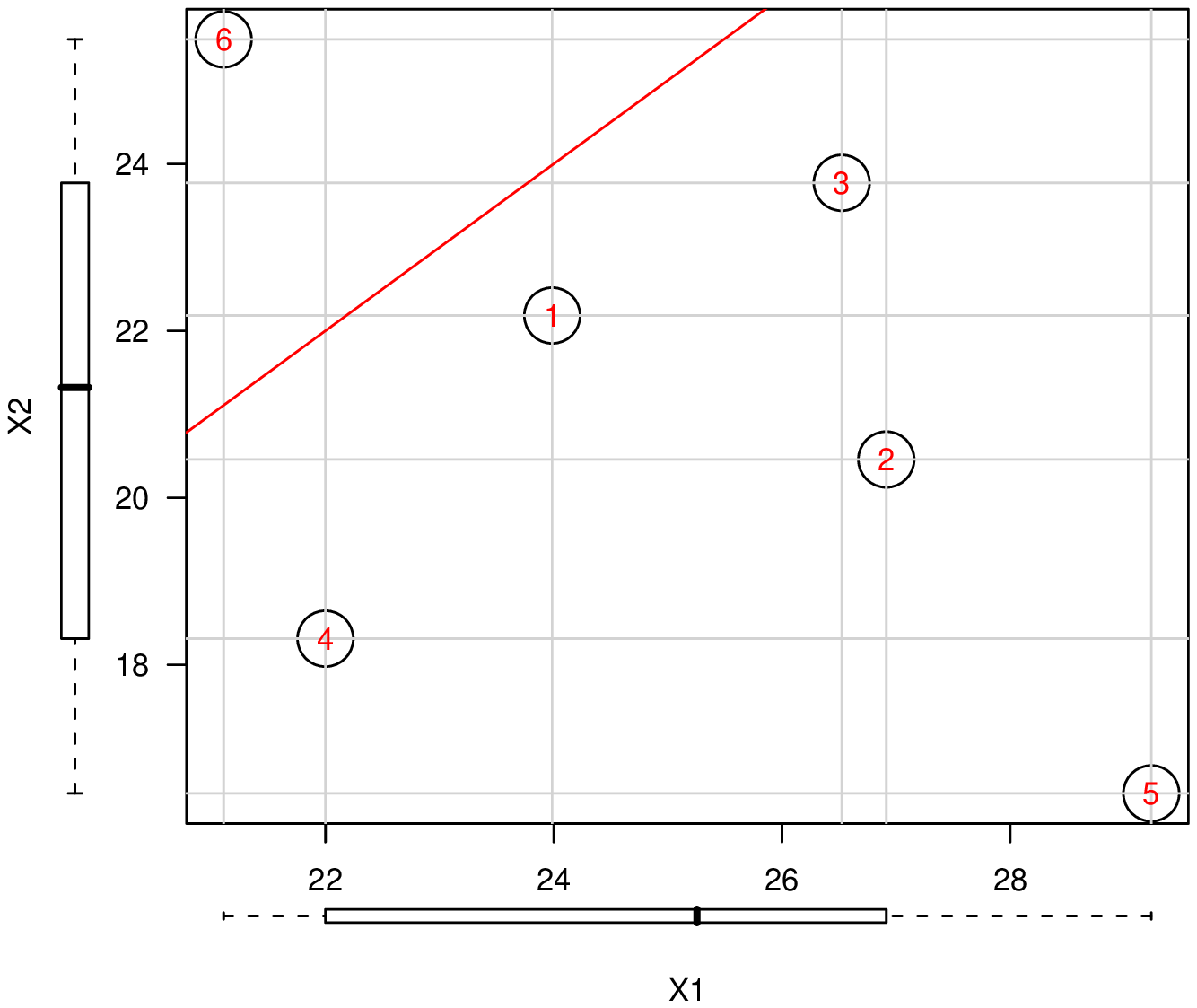,width=0.5\columnwidth}
\psfig{figure=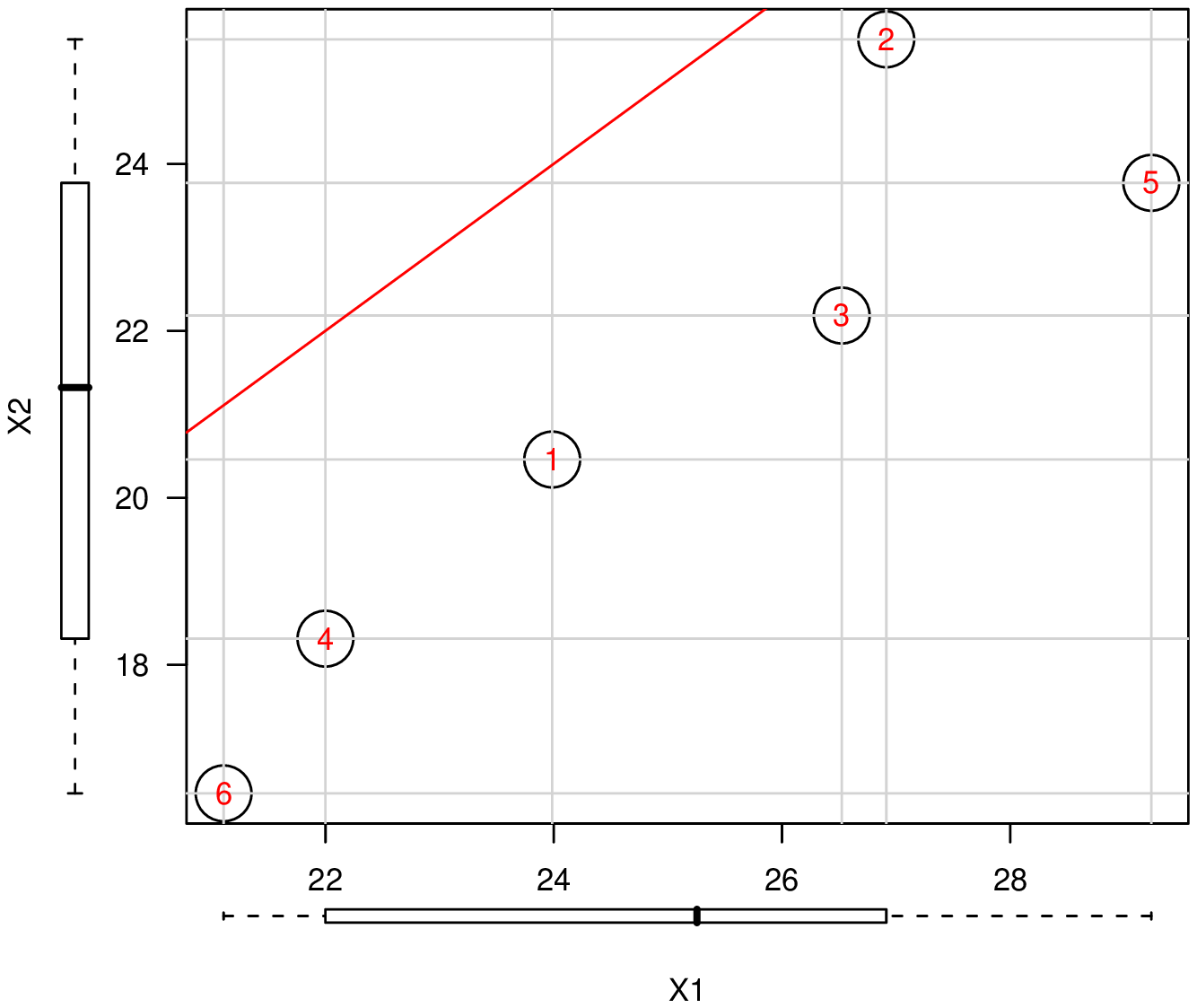,width=0.5\columnwidth}$$

\vspace{-0.8cm}
\centerline{(a) Initial LHS \hspace{3.5cm} (b) Final LHS}
\caption{Illustration of the cLHS algorithm ($n=6$) with a decreasing constraint between two uniformly-distributed random variables $X_1$ and $X_2$ ($\gamma(X_1,X_2)=18\%$).
The diagonal line corresponds to the frontier line $X_1=X_2$.}
  \label{fig:exdecroissant}
\end{figure}

We propose a simple example with a sample of size $n=6$ and two variables $X_1$ and $X_2$ subject to a decreasing constraint $X_1 > X_2$.
Points are uniformly sampled on the domain $\mathcal{X}=[20,30] \times [16,26]$ of $(X_1,X_2)$. 
We simulate an initial
 LHS (Figure~\ref{fig:exdecroissant}~(a)) and obtain the following design matrices:
\[
 X_1^n =
 \begin{pmatrix} 23.98\\26.91\\26.52\\21.99\\29.23\\21.10
 \end{pmatrix},\quad
 X_2^n =
 \begin{pmatrix} 22.18\\20.45\\23.77\\18.31\\16.45\\25.49
 \end{pmatrix},\quad
 \textrm{then } 
 \mathbf{C^n}=
 \begin{pmatrix}
 \mathbf{1} &1 &1 &0 &1 &0 \\
 1 &\mathbf{1} &1 &1 &1 &1 \\
 1 &1 &\mathbf{1} &0 &1 &0 \\
 1 &1 &1 &\mathbf{1} &1 &1 \\
 1 &1 &1 &1 &\mathbf{1} &1 \\
 0 &1 &1 &0 &1 &\mathbf{0} \\
 \end{pmatrix} .
\]
As the diagonal of this matrix has a null term, the equation (\ref{eq:objective}) is not fulfilled.
Then, the cLHS algorithm has to be applied in order to obtain a LHS satisfying the constraint.

First, we test the existence criterion.
We obtain $\mathbf{S}^n = (4 \; 6 \; 4 \; 6 \; 6 \; 3)$ and $\tilde{\mathbf{S}}^n = (3 \; 4 \; 4 \; 6 \; 6 \; 6)$.
The existence criterion (Eq. (\ref{eq:criterion})) is then fulfilled.

Second, we apply the algorithm and obtain the following result:
\begin{equation}
 X_2^{'n} =
 \begin{pmatrix} 20.45\\25.49\\22.18\\18.31\\23.77\\16.45
 \end{pmatrix}.
\end{equation}
$X_1^n$ has not been modified while elements of $X_2^n$ have been permuted to obtain $X_2^{'n}$, which is a sample satisfying the decreasing constraint.
Other $X_2^{'n}$ samples could be found, the choice made during the cLHS algorithm being random.
Figure~\ref{fig:exdecroissant}~(b) shows our final sampling result.

\section{An application case: welding thermomechanical models}

The robust increase in computer power has tremendously contributed to a growing fad for welding simulation. 
The industrial requirements are more and more numerous: supports to develop new processes, control of mechanical welding effects (in particular, residual stresses and distortions), argument in a nuclear safety analysis reports, etc. Thus, through the use of high-performance computers and advanced models, numerical simulation is expected to become an important tool for innovation in welding engineering.

However, running a welding simulation model requires a large number of inputs - about $500$ - including for example meshing inputs, boundary and initial conditions as well as material properties and process parameters, and generates several outputs, including spatial distributions of displacements and residual stresses in the weldment. 
In particular, among inputs, the determination of material properties is one of the key problems of welding simulation. 
The features of material properties are that they are dependent on temperature and that their full characterization is very expensive, often difficult or even sometimes impossible. 
In this context, the global sensitivity analyses of the numerical welding simulation model allows to determine which material properties are the most sensitive in a numerical welding simulation and in which range of temperature \citep{pet07,petass06,asslor09}.

Let us show the application of our methodology on the range of steel material.
Five input variables are the mechanical properties used by the model: Young's modulus, thermal computation coefficient, Poisson's ratio, yield strength and hardening modulus.
Because of their dependence on temperature, it has been required to sample each material property at a discrete set of temperatures: $7$ levels are chosen from $20^o$C to $1100^o$C.
We obtain $35=5 \times 7$ input variables, each following an uniform distribution defined by its minimal and maximal bounds (taken from the literature). 
Moreover, some material properties used in this model are monotonically decreasing as function of temperature.
Therefore, the constrained Latin hypercube sampling strategy, described in this paper, can be used to generate the input design. 
This strategy allows our sampled variables to honor their uniform repartition defined by their minimal and maximal bounds
, that is to say sampling in the physical bounds.

\begin{figure}[!ht]
    \centering
        \psfig{figure=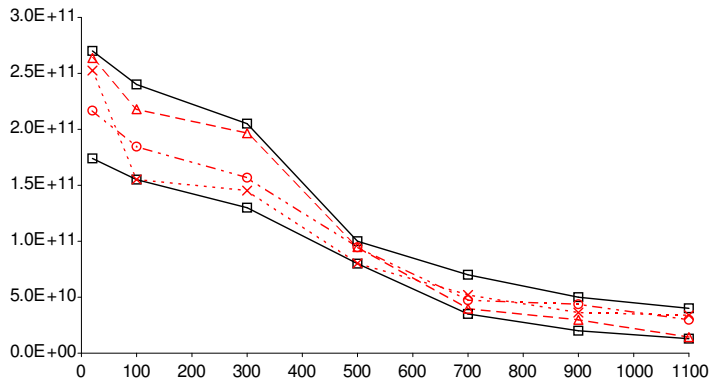,width=0.8\columnwidth}
    \caption{Example of $n=3$ samples of the Young's modulus obtained by cLHS (dashed lines). The upper and the lower curves (solid lines) represent the bounds of the domain for these $p=7$ variables.}
    \label{fig:5-dom-YOUN}
\end{figure}

To illustrate this application, we present the sampling of the Young's modulus.
Figure~\ref{fig:5-dom-YOUN} shows the result on this 7-dimensional variable which follows a decreasing constraint. 
These curves present three randomly selected materials among the $800$ created and the bounds of the domain. 
For the sensitivity analysis process, the Young's modulus is represented by only $7$ parameters \citep{asslor09}.
However, one should keep in mind that for the mechanical computation, the curve represents truly the considered dependence of this modulus because the algorithm uses intermediate values according to a piecewise linear interpolation.

\vspace{-0.5cm}

\vspace{-0.5cm}
It would be interesting to present the sensitivity analysis results obtained by employing the classical LHS (unconstrained) and by employing the constrained LHS. However some model calculations do not converge when using non physical evolution of the material properties (i.e. unconstrained LHS). 
Moreover, performing model calculations with non monotonic evolutions would be an aberration from a physical point of view.



\section{Conclusion}

In this paper, we have proposed a new algorithm (called cLHS) allowing to obtain samples of several variables constrained by some inequality relations.
This situation can frequently arise in application cases while very few works have been devoted to this issue.
The cLHS algorithm allows to satisfy the inequality constraint while leaving unchanged the one-dimensional marginals that we have defined for each variable.
In order to honor these one-dimensional marginals, the LHS-based technique has been preferred.
To our knowledge, this inequality constraint problem has not been studied for the LHS building issue.

We have shown the interest of this algorithm in an application case involving welding simulation model.
The cLHS algorithm has proven its efficiency to sample a multi-dimensional variable taking under consideration its physical nature.

The current cLHS algorithm has one main drawback.
When the minimal (respectively maximal) bounds of the two constrained variables move closer, the space filling properties of the sample points deteriorate: the sample points are gathered along the inequality frontier line.
The cLHS is therefore efficient if the bounds between the variables are sufficiently distant.
A constraint intensity measurement, noticed $\gamma$, has been defined in order to quantify this effect.
Moreover, a linear relation has been proposed between $\gamma$ and the correlation coefficient of the constrained variables.
This allows to a priori know (from the variable bound values), the effects of the inequality constraint in terms of correlation of the simulated sample. 

In a future work, it will be interesting to quantify this phenomenon by linking a space filling measure (as the discrepancy) with the bound values.
More generally, if we want to suppress this undesired effect, the LHS framework has to be left out.
A first idea would be to work with entropy-based designs by optimizing the entropy on one-dimensional marginals.
Developing design optimization algorithms under inequality constraints would be an interesting research way.

If more than two variables are under study, the cLHS algorithm is limited to inequality constraints defined in a sequential order.
A complex inequality constraint could involve more than two variables (e.g. $X_1<X_2+X_3$) or could be non sequential (e.g. $X_1<X_2$ and $X_1<X_3$).
If such situations are identified in specific application cases, there is no doubt that some extensions of the cLHS algorithm are possible.

\vspace{0.5cm}
{\it Acknowlegments}\;\; We thank O. Roustant and D. Ginsbourger, the organizers of the ENBIS-EMSE 2009 conference.
We are also grateful to the editors of this special issue
and anonymous referees for comments and suggestions.


\vspace{-0.5cm}
\bibliographystyle{apalike}

\end{document}